\documentclass{article}
\usepackage[T1]{fontenc}
\usepackage[utf8]{inputenc}
\usepackage{prettyref}
\usepackage{amstext}
\usepackage[numbers]{natbib}

\makeatletter
\usepackage{prettyref}
\newrefformat{eq}{eqn\nolinebreak\,\nolinebreak(\ref{#1})}
\newrefformat{sec}{Section\nolinebreak\,\nolinebreak\ref{#1}}
\newrefformat{cha}{Chapter\nolinebreak\,\nolinebreak\ref{#1}}
\newrefformat{tab}{Table\nolinebreak\,\nolinebreak\ref{#1}}
\newrefformat{fig}{Figure\nolinebreak\,\nolinebreak\ref{#1}}
\newrefformat{lem}{Lemma\nolinebreak\,\nolinebreak\ref{#1}}
\newrefformat{thm}{Theorem\nolinebreak\,\nolinebreak\ref{#1}}
\newrefformat{axm}{Axiom\nolinebreak\,\nolinebreak\ref{#1}}
\newrefformat{app}{Appendix\nolinebreak\,\nolinebreak\ref{#1}}
\newrefformat{exe}{Exercise\nolinebreak\,\nolinebreak\ref{#1}}
\newrefformat{cor}{Corollary\nolinebreak\,\nolinebreak\ref{#1}}
\newrefformat{pro}{Property\nolinebreak\,\nolinebreak\ref{#1}}
\newrefformat{def}{Definition\nolinebreak\,\nolinebreak\ref{#1}}
\newrefformat{subsec}{Section\nolinebreak\,\nolinebreak\ref{#1}}
\usepackage{amsmath}
\usepackage{bm}
\numberwithin{equation}{section}
\numberwithin{table}{section}
\newcommand{\br}[1]{\mathbf{#1}} 
\newcommand{\bg}[1]{\bm{#1}} 
\usepackage{pifont}
\usepackage{txfonts}
\usepackage{charter}
\date{}

\makeatother

\begin{document}

\title{Helmholtz Theorem and Uniqueness}

\author{{\normalsize{}Oliver Davis Johns}\\
{\normalsize{}San Francisco State University, Physics and Astronomy
Department}\\
{\normalsize{}1600 Holloway Avenue, San Francisco, CA 94133, USA}\\
{\normalsize{}Email: ojohns@metacosmos.org}\\
{\normalsize{}Web: http://www.metacosmos.org}}
\maketitle
\begin{abstract}
Vector calculus in $E^{3}$, three dimensions with a Euclidian metric,
is the \emph{lingua franca} of classical physics, including classical
electrodynamics. This article corrects some long-standing imprecisions
in a fundamental result.

Some textbooks assert that a vector function defined in the whole
of a three dimensional space is uniquely determined by its divergence,
its curl, and the condition that the function goes to zero as the
radius (distance from an origin) goes to infinity. This article suggests
that this condition is not sufficient for uniqueness. A proof is given
that a sufficient condition for uniqueness is for the vector function
to approach zero more rapidly than radius to the minus 3/2 power as
the radius goes to infinity.

The issue is important because the same uniqueness condition also
determines the uniqueness of the decomposition of a vector field into
a transverse field plus a solenoidal field, as is done in the Coulomb
gauge of electrodynamics. 
\end{abstract}

\section{Introduction\label{sec:Introduction}}

With the exception of the Dirac delta function in Appendix A, all
functions in this article are assumed to have continuous derivatives
to all orders. The statement that a field function $f(\br r)$ obeys
$f=O(r^{-\alpha})$ as $r\rightarrow\infty$, where $\alpha$ is a
given positive constant, is defined to mean that there exist a positive
constant $c$ and a radius $r_{0}$ such that $\left|f(\br r)\right|<cr^{-\alpha}$
for all $r>r_{0}.$ (Equivalently, that $r^{\alpha}\left|f(\br r)\right|$
is bounded as $r\rightarrow\infty$.) For vector fields $\br F(\br r)$
the statement that $\br F=O(r^{-\alpha})$ is defined to mean that,
for each cartesian component, $F_{i}=O(r^{-\alpha})$ where $i=[1,2,3].$ 

Some standard electromagnetism textbooks assert that a vector function
$\br F(\br r)$ defined over the whole space $E^{3}$ is determined
uniquely by its divergence, its curl, and the condition that $\br F$
vanishes at infinity. If that condition is interpreted as $\br F=O(r^{-\alpha})$
for any positive constant $\alpha$, then this article proves that
condition to be too weak to imply uniqueness. The sufficient condition
for uniqueness is proved here to be $\br F=O(r^{-(3/2+\beta)})$,
where $\beta$ is any (possibly small) positive constant. That is,
$\alpha$ must be greater than $3/2$; function $\br F$ must go to
zero more rapidly than $r^{-3/2}$.

Some textbook treatments of the uniqueness of a function given its
divergence and curl in the whole of $E^{3}$ are discussed in \prettyref{sec:Text-Treatments}.
Two older textbooks give a condition for uniqueness that is sufficient
but unnecessarily strong, and two more recent textbooks assert a condition
that is too weak and not sufficient.

\section{Basic Theorems\label{sec:Basic-Theorems}}

\textbf{Divergence Theorem:} Let $\br F(x,y,z)$ be defined in a volume
${\cal V}$ surrounded by a sufficiently smooth boundary ${\cal S}$.
Denote the three dimensional volume element by $d\tau=d^{3}r$ and
the outwards pointing surface element by $d\br a$. Then the divergence
$\left(\nabla\cdot\br F\right)$ is related to the surface integral
by
\begin{equation}
\int_{{\cal V}}\!\left(\nabla\cdot\br F\right)\,d\tau=\oint_{{\cal S}}\br F\cdot d\br a\label{eq:2.1}
\end{equation}
\textbf{Stokes Theorem:} Let $\br F(x,y,z)$ be defined on a surface
${\cal S}$ surrounded by a sufficiently smooth boundary line ${\cal C}$.
(This surface need not be planar.) Denote the line element of the
boundary by $d\bg\ell$ and the surface element of ${\cal S}$ by
$d\br a$. Assume $d\bg\ell$ to point in the direction of the fingers
of the right hand when its thumb points in the direction of $d\br a$.
Then the curl $\left(\nabla\times\br F\right)$ is related to the
line integral by 
\begin{equation}
\int_{{\cal S}}\!\left(\nabla\times\br F\right)\cdot d\br a=\oint_{{\cal C}}\br F\cdot d\bg\ell\label{eq:2.2}
\end{equation}

\section{Existence of Scalar Potential $\phi$\label{sec:Existence-of-Scalar}}

\textbf{Theorem 3.1:} These three statements are equivalent. Any one
of the following three items is true if and only if all three are
true.
\begin{enumerate}
\item $\nabla\times\br F=0$ in the whole of three-dimensional space $E^{3}$.
\item $\oint\br F\cdot d\bg\ell=0$ for any closed line integral of $\br F$
in the three dimensional space.
\item There exists a single-valued function $\phi(\br r)$ such that $\br F=-\nabla\phi$
at all points of the three-dimensional space.
\end{enumerate}
\textbf{Proof:} Since the path of the closed line integral is arbitrary,
the equivalence of Items 1 and 2 follows directly from Stokes Theorem. 

To prove the equivalence of Item 3, choose a path from the origin
to point $x,y,z$ and define $\phi$ as
\begin{equation}
\phi(\br r)=-\int_{0}^{\br r}\br F\cdot d\bg\ell=-\int_{0}^{x,y,z}\left(F_{x}dx+F_{y}dy+F_{z}dz\right)\label{eq:4.1}
\end{equation}
Then 
\begin{equation}
\dfrac{\partial\phi}{\partial x}=-\lim_{\delta\rightarrow0}\dfrac{1}{\delta}\left(\int_{0}^{x+\delta,y,z}-\int_{0}^{x,y,z}\right)\left(F_{x}dx+F_{y}dy+F_{z}dz\right)=-F_{x}\label{eq:4.2}
\end{equation}
with similar results for $y$ and $z$, and hence
\begin{equation}
\br F(\br r)=-\nabla\phi(\br r)\label{eq:4.3}
\end{equation}

To prove $\phi$ single valued, note that any two \emph{different}
paths from the origin to $x,y,z$ can be combined to form a closed
path. This combined path will go from the origin to $\br r$ on one
path and then, in the reverse sense, from $\br r$ back to the origin
on the other path. But integration in a reverse sense on a path simply
changes the sign of that integral. From item 2 above, the combined
path must yield zero, and hence those two different paths from the
origin to $x,y,z$ must yield the same integral. The definition of
$\phi$ in \prettyref{eq:4.1} is path independent and $\phi$ is
a well-defined, single-valued function.

Thus Item 2 implies Item 3. And, from the curl of \prettyref{eq:4.3},
Item 3 implies Item 1. But Items 1 and 2 are equivalent. Thus all
three statements are equivalent. $\blacksquare$

\section{Alternate Definition of Scalar Potential\label{sec:Alternate-Definition-of}}

Although the definition of $\phi$ in \prettyref{eq:4.1} is adequate,
it is not unique. If $\nabla\phi=-\br F$ then it is also true that
$\nabla(\phi+b)=-\br F$, where $b$ is any constant. It is useful
to choose $b$ to cancel the lower limit in \prettyref{eq:4.1} and
 define an alternate single valued function $\tilde{\phi}$ as
\begin{equation}
\tilde{\phi}(\br r)=\phi(\br r)+b=-\int^{\br r}\br F\cdot d\bg\ell\label{eq:4.4}
\end{equation}
Of course, since
\begin{equation}
-\nabla\tilde{\phi}=-\nabla(\phi+b)=-\nabla\phi=\br F\label{eq:4.5}
\end{equation}
the alternate function still satisfies \prettyref{eq:4.3}. 

If $\alpha>1$, it follows from \prettyref{eq:4.4} that the behaviors
of $\br F$ and $\tilde{\phi}$ as $r\rightarrow\infty$ are related
by
\begin{equation}
\br F=O(r^{-\alpha})\iff\tilde{\phi}=O(r^{-\alpha+1})\label{eq:4.6}
\end{equation}

\section{Uniqueness of a Vector Field \protect \\
in a Finite Volume\label{sec:Uniq-finite} }

A function $\br F(\br r)$ is defined in a finite volume ${\cal V}$
with surface ${\cal S}$. From this function we can calculate its
divergence $s(\br r)$ and curl $\br C(\br r)$
\begin{equation}
\nabla\cdot F=s\quad\quad\quad\quad\nabla\times\br F=\br C\label{eq:5.1}
\end{equation}
Denoting the outward pointing surface element of ${\cal S}$ as $d\br a=\hat{\br n}\,da$,
we can also calculate the outward normal 
\begin{equation}
F_{n}=\hat{\br n}\cdot\br F\label{eq:5.1a}
\end{equation}

\textbf{Theorem 5.1:} A function $\br F$ defined in a finite volume
${\cal V}$ with surface ${\cal S}$ is uniquely determined by its
divergence and curl in ${\cal V}$, and its outward normal on ${\cal S}$,
in the sense that any other function that has these same values must
be equal to $\br F$.

\textbf{Proof:} Suppose two functions $\br F^{*}$and $\br F$  both
to have the divergence, curl, and outward normal given in \prettyref{eq:5.1}
and \prettyref{eq:5.1a}. The proof is to define 
\begin{equation}
\br W=\br F^{*}-\br F\label{eq:5.2}
\end{equation}
and then prove $\br W=0$. 

This definition of $\br W$ implies that
\begin{equation}
\nabla\cdot\br W=0\quad\quad\quad\nabla\times\br W=0\quad\quad\quad W_{n}=0\label{eq:5.3}
\end{equation}
Since $\nabla\times\br W=0$ it follows from \prettyref{sec:Existence-of-Scalar}
with $\br W$ in place of $\br F$ that there exists a scalar function
$\phi$ such that 
\begin{equation}
\br W=-\nabla\phi\label{eq:5.4}
\end{equation}
 Since $\nabla\cdot\br W=0$, this $\phi$ obeys 
\begin{equation}
\nabla^{2}\phi=-\nabla\cdot\br W=0\label{eq:5.5}
\end{equation}
Expanding the divergence of $\left(\phi\nabla\phi\right)$ and using
\prettyref{eq:5.4} and \prettyref{eq:5.5} gives 
\begin{equation}
\nabla\cdot\left(\phi\nabla\phi\right)=\phi\left(\nabla^{2}\phi\right)+\left(\nabla\phi\right)\cdot\left(\nabla\phi\right)=\left(\nabla\phi\right)\cdot\left(\nabla\phi\right)\label{eq:5.6}
\end{equation}
and hence
\begin{equation}
\nabla\cdot\left(\phi\nabla\phi\right)=\left(\nabla\phi\right)\cdot\left(\nabla\phi\right)=W^{2}\label{eq:5.7}
\end{equation}
Now apply the divergence theorem to give 
\begin{equation}
\int_{{\cal V}}W^{2}d\tau=\int_{{\cal V}}\nabla\cdot\left(\phi\nabla\phi\right)d\tau=\oint_{{\cal S}}\left(\phi\nabla\phi\right)\cdot d\br a=-\oint_{{\cal S}}\left(\phi W_{n}da\right)\label{eq:5.8}
\end{equation}
Since $W_{n}=\left(F_{n}^{*}-F_{n}\right)=0$, and $W^{2}$ is positive
definite, it follows that $\br W=0$ and hence $\br F^{*}=\br F$,
which proves uniqueness in ${\cal V}$. $\blacksquare$

\textbf{Caution: }It is tempting to apply Theorem 5.1 by assuming
condition $\br F\rightarrow0$ as $r\rightarrow\infty$ to mean that
$W_{n}=0$ on a surface ${\cal S}$ \emph{at infinity, }which would
prove uniqueness. But there is no such thing as a surface \emph{at
infinity.} Infinity is a limit, not a location. When the volume ${\cal V}$
expands to include the whole of $E^{3}$, proof of uniqueness requires
a limiting process as surface ${\cal S}$ \emph{goes to infinity,}
which it never reaches. A correct proof of uniqueness in the whole
of $E^{3}$ is given in \prettyref{sec:Uniq-whole}.

\section{Uniqueness of a Vector Field \protect \\
in the Whole of $E^{3}$\label{sec:Uniq-whole}}

If ${\cal V}$ is extended to be the whole space, the uniqueness condition
$W_{n}=0$ in \prettyref{eq:5.8} can be replaced by the condition
that $\br W$ goes to zero sufficiently rapidly as $r\rightarrow\infty$. 

\textbf{Theorem 6.1: }A function $\br F$ defined on all space is
uniquely determined by its divergence, its curl, and the condition
that, for some (possibly small) positive constant $\beta$, $\br F=O(r^{-(3/2+\beta)})$
as $r\rightarrow\infty$, in the sense that any other function that
meets these same conditions must be equal to $\br F$.

\textbf{Proof:} As in \prettyref{sec:Uniq-finite}, assume that functions
$\br F^{*}$ and $\br F$ have the same divergence and curl and define
$\br W=\left(\br F^{*}-\br F\right)$. We prove that $\br W=0$, which
implies $\br F^{*}=\br F$ and hence uniqueness. 

It follows from the definition of $\br W$ that $\nabla\cdot\br W=0$
and $\nabla\times\br W=0$. Theorem 3.1 with $\br W$ in place of
$\br F$ then permits the definition of a scalar function $\phi$
with $\br W=-\nabla\phi$. The \prettyref{eq:4.4} of \prettyref{sec:Alternate-Definition-of}
can then be used to define an alternate function $\tilde{\phi}$,
such that 
\begin{equation}
\br W=-\nabla\tilde{\phi}\quad\quad\text{and hence}\quad\quad\nabla^{2}\tilde{\phi}=\nabla\cdot\br W=0\label{eq:20.0}
\end{equation}
Expanding the divergence of $\left(\tilde{\phi}\,\nabla\tilde{\phi}\right)$
and using \prettyref{eq:20.0} gives
\begin{equation}
\nabla\cdot\left(\tilde{\phi}\,\nabla\tilde{\phi}\right)=\tilde{\phi}\left(\nabla^{2}\tilde{\phi}\right)+\left(\nabla\tilde{\phi}\right)\cdot\left(\nabla\tilde{\phi}\right)=\left(\nabla\tilde{\phi}\right)\cdot\left(\nabla\tilde{\phi}\right)\label{eq:20.1}
\end{equation}
and hence 
\begin{equation}
\nabla\cdot\left(\tilde{\phi}\,\nabla\tilde{\phi}\right)=\left(\nabla\tilde{\phi}\right)\cdot\left(\nabla\tilde{\phi}\right)=W^{2}\label{eq:20.2}
\end{equation}
Applying the divergence theorem for a volume ${\cal V}_{\rho}$ enclosed
by a spherical surface ${\cal S}_{\rho}$ of radius $\rho$ centered
on the origin gives
\begin{equation}
\int_{{\cal V}_{\rho}}W^{2}d\tau=\int_{{\cal V}_{\rho}}\nabla\cdot\left(\tilde{\phi}\,\nabla\tilde{\phi}\right)d\tau=\oint_{{\cal S}_{\rho}}\left(\tilde{\phi}\,\nabla\tilde{\phi}\right)\cdot d\br a\label{eq:20.3}
\end{equation}
As $\rho\rightarrow\infty$, this ${\cal V}_{\rho}$ will expand to
become the whole space.

Suppose that $\br W=-\nabla\tilde{\phi}=O(r^{-\alpha})$ as $r\rightarrow\infty$,
for some $\alpha>1$. From \prettyref{eq:4.6} it follows that $\tilde{\phi}=O(r^{-\alpha+1})$
and hence  $\left(\tilde{\phi}\,\nabla\tilde{\phi}\right)=O(r^{-2\alpha+1})$.
Taking account of the fact that $da=r^{2}d\Omega$ for an increment
of solid angle $d\Omega$, the surface integral in \prettyref{eq:20.3}
is then $O(r^{-2\alpha+3})$ and will approach zero as $r\rightarrow\infty$
provided that $-2\alpha+3<0$ and hence $\alpha>3/2$. If the two
functions $\br F^{*}$ and $\br F$, and thus their difference $\br W$,
obey $O(r^{-(3/2+\beta)})$, for some positive constant $\beta$,
the surface integral in \prettyref{eq:20.3} vanishes as $\rho\rightarrow\infty$
and ${\cal V}_{\rho}$ becomes the whole of $E^{3}$. Since $W^{2}$
is positive definite, it then follows from \prettyref{eq:20.3} that
$\br W=0$ and hence $\br F^{*}=\br F$, which proves uniqueness.
The condition for uniqueness can also be stated as $\tilde{\phi}=O(r^{-(1/2+\beta)})$.
(As a check, notice that $\alpha=3/2+\beta>1$ as was assumed.) $\blacksquare$

\textbf{Note 6.1:} It follows from Theorem 6.1 that any vector function
which has zero divergence, zero curl, and goes to zero faster than
$r^{-3/2}$ as $r\rightarrow\infty$ must be identically zero. There
is no (nonzero) function that has zero divergence, zero curl, and
goes to zero faster than $r^{-3/2}$ as $r\rightarrow\infty.$

\section{Solution of the Poisson Equation in $E^{3}$\label{sec:Solution-Poisson}}

\textbf{Theorem 7.1:} Given a source function $f(\br r),$ the Poisson
equation for function $U(\br r)$ is (with the conventional minus
sign)
\begin{equation}
\nabla^{2}U(\br r)=-f(\br r)\label{eq:6.1}
\end{equation}
If $f=O(r^{-(2+\varepsilon)})$ as $r\rightarrow\infty$, where $\varepsilon$
is some positive constant, then a solution of the Poisson Equation
at point $\br r_{0}$ is
\begin{equation}
U(\br r_{0})=\dfrac{1}{4\pi}\int\dfrac{f(\br r)}{\left|\br r-\br r_{0}\right|}d\tau\label{eq:6.2}
\end{equation}
\textbf{Proof:} Spherical polar coordinates can be introduced in \prettyref{eq:6.2}
without loss of generality. Then $d\tau=r{}^{2}\,d\Omega\,dr$ and
the integrand in \prettyref{eq:6.2} is $O(r^{-\left(2+\varepsilon\right)}r{}^{-1}r^{2})$
which is $O(r^{-(1+\varepsilon)})$, and hence the integral defining
$U(\br r_{0})$ converges. 

As shown in Appendix A, the condition $f=O(r^{-(2+\varepsilon)})$
is also sufficient for the use of Green's identity to establish the
correctness of \prettyref{eq:6.2}. $\blacksquare$

\textbf{Note 7.1:} When, as here, the Poisson Equation is assumed
to hold in the whole of space $E^{3}$, with only the condition $f=O(r^{-\left(2+\varepsilon\right)})$
to ensure the convergence of \prettyref{eq:6.2}, then its solution
$U(\br r)$ is not uniquely determined. For if $U$ is a solution
then $\tilde{U}=(U+b)$ is also a solution, where $b$ is any constant.

However, even though $U$ itself is not uniquely determined, its gradient
may be. If $\nabla U$ obeys the condition $\nabla U=O(r^{-(3/2+\beta)})$,
then Theorem 6.1 proves that $\nabla U$ is uniquely determined by
its divergence $\nabla\cdot(\nabla U)=-f$ and curl $\nabla\times\nabla U=0.$ 

\textbf{Note 7.2:} If $f$ is nonzero only at finite distance from
the origin (a more stringent condition than $f=O(r^{-(2+\varepsilon)})$),
then \prettyref{eq:6.2} implies that $\nabla U=O(r^{-2})$. Since
$2>3/2$, the vector $\nabla U$ will then be uniquely determined.
For example, putting $f$ and $-\nabla U$ equal to charge density
and electric field, respectively, the electric field of a static,
spatially finite charge distribution is uniquely determined.

\section{Unique Solution of Poisson Equation \protect \\
in Finite Volumes\label{sec:Uni-Poisson-Finite}}

Computation of $U$ is often aided by uniqueness theorems involving
boundary conditions, specification of $U$ on the boundary surface
${\cal S}$ of a volume ${\cal V}$.

Let $U^{*}$ and $U$ be any two potentially different solutions to
\prettyref{eq:6.1} and define $\eta=(U^{*}-U)$. Any boundary condition
sufficient to make $\eta=0$ in the whole of ${\cal V}$ will imply
uniqueness, in the sense that there can be only one solution in ${\cal V}$
satisfying that boundary condition.

Since we have both \prettyref{eq:6.1} and $\nabla^{2}U^{*}(\br r)=-f(\br r)$
it follows that $\nabla^{2}\eta=0$. Following the pattern in \prettyref{sec:Uniq-finite}
gives
\begin{equation}
\nabla\cdot(\eta\nabla\eta)=\eta\left(\nabla^{2}\eta\right)+\left(\nabla\eta\right)\cdot\left(\nabla\eta\right)=\left(\nabla\eta\right)\cdot\left(\nabla\eta\right)\label{eq:30.2}
\end{equation}
\begin{equation}
\int_{{\cal V}}\left|\nabla\eta\right|^{2}d\tau=\int_{{\cal V}}\left(\nabla\eta\right)\cdot\left(\nabla\eta\right)d\tau=\int_{{\cal V}}\nabla\cdot(\eta\nabla\eta)=\int_{{\cal S}}\eta\nabla\eta\cdot d\text{a}\label{eq:30.3}
\end{equation}
We now use \prettyref{eq:30.3} to discuss two classes of boundary
conditions, Dirichlet and Neumann.

1. Dirichlet Boundary Condition: Suppose $U$ to be specified on ${\cal S}$
so that $U^{*}=U$ there. Then $\eta=0$ on ${\cal S}$, and the vanishing
of the surface integral in \prettyref{eq:30.3} shows that the positive
definite quantity $\left|\nabla\eta\right|$ is zero in the whole
of ${\cal V}$. The vanishing of gradient $\nabla\eta=0$ everywhere
in ${\cal V}$, together with $\eta=0$ on ${\cal S}$, imply $\eta=0$
for the whole of ${\cal V}$. Thus the two possibly different solutions
$U^{*}$and $U$ are equal. There is only one unique solution in ${\cal V}$
that has the specified boundary value on ${\cal S}$. 

For the particular example of the Laplace equation (\prettyref{eq:6.1}
with $f=0)$, an obvious possible solution is $U=0$. A Dirichlet
boundary condition that $U=0$ on ${\cal S}$ proves that solution
to be unique. For the Laplace equation, the only solution that vanishes
on ${\cal S}$ is the solution that also vanishes in the whole of
${\cal V}$.

2. Neumann Boundary Condition: Suppose $\nabla U$ to be given on
${\cal S}$ so that $\nabla U^{*}=\nabla U$ there. Then $\nabla\eta=0$
on ${\cal S}$, and the vanishing of the surface integral in \prettyref{eq:30.3}
implies that the positive definite quantity $\left|\nabla\eta\right|$
is zero in the whole of ${\cal V}$. The vanishing of gradient $\nabla\eta=0$
everywhere in ${\cal V}$ implies $\nabla U^{*}=\nabla U$ for the
whole of ${\cal V}$. There is only one unique gradient $\nabla U$
in ${\cal V}$ that has the specified boundary value $\nabla U$ on
${\cal S}$. 

Also, the result that $\left|\nabla\eta\right|=0$ in the whole of
${\cal V}$ implies that $\eta$ must be some constant $b$, and hence
$U^{*}=U+b$. For the Neumann boundary condition, $\nabla U$ is uniquely
determined and $U$ is uniquely determined up to an additive constant.

\textbf{Caution:} These Dirichlet and Neumann boundary results cannot
be applied directly to a surface or portion of a surface \emph{at
infinity}. As noted at the end of \prettyref{sec:Uniq-finite}, there
is no such thing as a surface \emph{at infinity.} Infinity is only
approached as a limit, as is done in \prettyref{sec:Uniq-whole}.

For example, if a solution to the Laplace equation goes to zero as
$r\rightarrow\infty$ (\emph{i.e.,} is $O(r^{-\varepsilon})$ for
some small $\varepsilon>0$) then one might invoke a Dirichlet boundary
condition that $U$ is zero on the surface ${\cal S}$ \emph{at infinity}
to conclude from \prettyref{eq:30.3} that such a function must be
zero for the whole of $E^{3}$. But that would be an incorrect assertion.
On must instead define a spherical surface of radius $\rho$ surrounding
a volume ${\cal V}$ and consider the vanishing of the surface integral
in \prettyref{eq:30.3} in the limit $\rho\rightarrow\infty$ when
${\cal V}$ becomes the whole of $E^{3}$. 

\section{Helmholtz Theorem\label{sec:Helmholtz-Theorem}}

\prettyref{sec:Uniq-whole} proved that there can be only one function
that vanishes sufficiently rapidly at infinity and has a particular
divergence and curl. But we still haven't proved that, given only
a proposed divergence and curl, we can actually construct  a function
with those properties. This is the task of the Helmholtz Theorem.

\textbf{Helmholtz Theorem: }Suppose a scalar function $s(\br r)$
and a divergence-less vector function $\br C(\br r)$ both of which
are $O(r^{-\left(2+\varepsilon\right)})$ as $r\rightarrow\infty$,
where $\varepsilon$ is some positive constant. Then there exist functions
$\phi(\br r)$ and $\br A(\br r)$ such that 
\begin{equation}
\br F(\br r)=-\nabla\phi+\nabla\times\br A\label{eq:7.1}
\end{equation}
defines a vector field $\br F$ with 
\begin{equation}
\nabla\cdot\br F=s\quad\quad\quad\quad\nabla\times\br F=\br C\label{eq:7.2}
\end{equation}

\textbf{Proof:} Since \textbf{$s=O(r^{-\left(2+\varepsilon\right)})$,
}Theorem 7.1 with $f=s$ and $U=\phi$ shows that the integral 
\begin{equation}
\phi(\br r_{0})=\dfrac{1}{4\pi}\int\dfrac{s(\br r)}{\left|\br r-\br r_{0}\right|}d\tau\label{eq:7.3}
\end{equation}
is convergent and defines a function $\phi$ such that
\begin{equation}
\nabla\cdot\left(-\nabla\phi(\br r)\right)=-\nabla^{2}\phi(\br r)=s(\br r)\label{eq:7.4}
\end{equation}

Since $\br C=O(r^{-\left(2+\varepsilon\right)})$, Theorem 7.1 with
$f=C_{i}$ and $U=A_{i}$ shows that the integral 
\begin{equation}
\br A(\br r_{0})=\dfrac{1}{4\pi}\int\dfrac{\br C(\br r)}{\left|\br r-\br r_{0}\right|}d\tau\label{eq:7.5}
\end{equation}
is convergent, and  defines vector function $\br A(\br r)$ such that
\begin{equation}
\nabla\cdot\left(-\nabla\br A(\br r)\right)=-\nabla^{2}\br A(\br r)=\br C(\br r)\label{eq:7.6a}
\end{equation}
Expansion of a double cross product then gives
\begin{equation}
\nabla\times\left(\nabla\times\br A\right)=-\nabla^{2}\br A+\nabla\left(\nabla\cdot\br A\right)=\br C+\nabla\left(\nabla\cdot\br A\right)\label{eq:7.6}
\end{equation}

To complete the proof, we now show that the vector $\br A$ defined
in \prettyref{eq:7.5} obeys $\left(\nabla\cdot\br A\right)=0$ and
hence the second term on the far right in \prettyref{eq:7.6} is zero.
From \prettyref{eq:7.5}, 
\begin{equation}
4\pi\nabla_{0}\cdot\br A(\br r_{0})=\int\br C(\br r)\cdot\nabla_{0}\dfrac{1}{\left|\br r-\br r_{0}\right|}\,d\tau=-\int\br C(\br r)\cdot\nabla\dfrac{1}{\left|\br r-\br r_{0}\right|}\,d\tau\label{eq:7.7}
\end{equation}
\begin{equation}
=\int\dfrac{1}{\left|\br r-\br r_{0}\right|}\,\,\left(\nabla\cdot\br C(\br r)\right)\,\,d\tau-\int\nabla\cdot\left(\dfrac{\br C(\br r)}{\left|\br r-\br r_{0}\right|}\right)d\tau\label{eq:7.8}
\end{equation}
The assumption that $\br C$ is divergence-less gives $\left(\nabla\cdot\br C(\br r)\right)=0$
which shows that the first integral in \prettyref{eq:7.8} is zero. 

Define a spherical surface ${\cal S}$ of radius $\rho$, centered
at the origin and enclosing a volume ${\cal V}$. As $\rho\rightarrow\infty$,
volume ${\cal V}$ will expand to become the whole space. For $\rho>r_{0}$,
the divergence theorem allows the second integral in \prettyref{eq:7.8}
to be written as a surface integral over ${\cal S}$.
\begin{equation}
\int_{{\cal V}}\nabla\cdot\left(\dfrac{\br C(\br r)}{\left|\br r-\br r_{0}\right|}\right)d\tau=\oint_{{\cal S}}\left(\dfrac{\br C(\br r)}{\left|\br r-\br r_{0}\right|}\right)\cdot d\br a\label{eq:7.9}
\end{equation}
The surface ${\cal S}$ has $da=r^{2}d\Omega$ for increments of solid
angle $d\Omega$, where $r=\rho$. The surface integral in \prettyref{eq:7.9}
is then  $O(r^{-2-\varepsilon-1+2})=O(r{}^{-1-\varepsilon})$ and
hence vanishes as $\rho\rightarrow\infty$ and ${\cal S}$ expands
to enclose the whole space. Thus $4\pi\nabla_{0}\cdot\br A(\br r_{0})=0$
as required. The \prettyref{eq:7.6} then becomes
\begin{equation}
\nabla\times\left(\nabla\times\br A\right)=-\nabla^{2}\br A=\br C\label{eq:7.10}
\end{equation}

Taking the divergence of \prettyref{eq:7.1} and using \prettyref{eq:7.4}
gives
\begin{equation}
\nabla\cdot\br F=-\nabla^{2}\phi+\nabla\cdot\left(\nabla\times\br A\right)=s\label{eq:7.11}
\end{equation}
the first of \prettyref{eq:7.2}. Taking the curl of \prettyref{eq:7.1}
and using \prettyref{eq:7.10} gives 
\begin{equation}
\nabla\times\br F=-\nabla\times\left(\nabla\phi\right)+\nabla\times\left(\nabla\times\br A\right)=\br C\label{eq:7.12}
\end{equation}
the second of \prettyref{eq:7.2}, which completes the proof. $\blacksquare$

\textbf{Note 9.1:} The potential functions $\phi$ and $\br A$ are
not unique. A function $\tilde{\phi}=\phi+b$, where $b$ is an arbitrary
constant, would have the same gradient. Also a function $\tilde{\br A}=\br A+\nabla\theta$
where $\theta$ is an arbitrary field function, would have the same
curl. Also, such an alternate function $\tilde{\br A}$ would have
$\nabla\cdot\tilde{\br A}=\nabla^{2}\theta$ which need not be zero.

\section{Decomposition of a Vector Function\label{sec:Decomposition}}

\textbf{Theorem 10.1: }Consider any vector field $\br F(\br r)$.
By calculation, this field has divergence and curl 
\begin{equation}
\nabla\cdot\br F=s\quad\quad\quad\quad\nabla\times\br F=\br C\label{eq:8.1}
\end{equation}
If both $s$ and $\br C$ are $O(r^{-\left(2+\varepsilon\right)})$
as $r\rightarrow\infty,$ then there exist a function $\br F_{T}$
called the \emph{transverse field} and a function $\br F_{S}$ called
the \emph{solenoidal field,} such that $\br F$ can be decomposed
into the sum
\begin{equation}
\br F=\br F_{T}+\br F_{S}\label{eq:8.2}
\end{equation}
where 
\begin{equation}
\nabla\cdot\br F_{T}=s\quad\quad\text{and}\quad\quad\nabla\times\br F_{T}=0\label{eq:8.3}
\end{equation}
and
\begin{equation}
\nabla\cdot\br F_{S}=0\quad\quad\text{and}\quad\quad\nabla\times\br F_{S}=\br C\label{eq:8.4}
\end{equation}
If both $\br F_{T}$ and $\br F_{S}$ vanish as $O(r^{-(3/2+\beta)})$
as $r\rightarrow\infty$, then they are uniquely determined and the
decomposition in \prettyref{eq:8.2} is unique.

\textbf{Proof:} Since the functions $s$ and $\br C$ are both $O(r^{-\left(2+\varepsilon\right)})$,
the Helmholtz Theorem of \prettyref{sec:Helmholtz-Theorem} proves
the existence of potential functions $\phi$ and $\br A$ such that
\begin{equation}
\br F=-\nabla\phi+\nabla\times\br A\label{eq:8.5}
\end{equation}
 satisfies \prettyref{eq:8.1}. 

It follows that the functions defined as 
\begin{equation}
\br F_{T}=-\nabla\phi\quad\quad\text{and}\quad\quad\br F_{S}=\nabla\times\br A\label{eq:8.6}
\end{equation}
satisfy \prettyref{eq:8.2}, \prettyref{eq:8.3},  and \prettyref{eq:8.4},
as was to be proved.

Theorem 6.1 shows that if $\br F_{T}=O(r^{-(3/2+\beta)})$ then $\br F_{T}$
is uniquely determined by \prettyref{eq:8.3}. Also, if $\br F_{S}=O(r^{-(3/2+\beta)})$
then $\br F_{S}$ is uniquely determined by \prettyref{eq:8.4}. With
both $\br F_{T}$ and $\br F_{S}$ uniquely determined, the decomposition
in \prettyref{eq:8.2} is unique. $\blacksquare$

\textbf{Note 10.1:} The discussion at the end of \prettyref{sec:Solution-Poisson}
applied with $f$ replaced by $s$ shows that, if $s$ is nonzero
only at finite distance from the origin (a more stringent condition
than $s=O(r^{-(2+\varepsilon)})$), then \prettyref{eq:6.2} implies
that $\nabla\phi=O(r^{-2})$. Since $2>3/2$, the vector $\br F_{T}=-\nabla\phi$
will then be uniquely determined. Similarly, if $\br C$ is nonzero
only at finite distance from the origin (a more stringent condition
than $\br C=O(r^{-(2+\varepsilon)})$), then \prettyref{eq:6.2} implies
that $\nabla\times\br A=O(r^{-2})$. Since $2>3/2$, the vector $\br F_{S}=\nabla\times\br A$
will then be uniquely determined.

\textbf{Note 10.2:} Let $\br F=\br F_{T}+\br F_{S}$ and $\br G=\br G_{T}+\br G_{S}$,
where $\br F_{T}$, $\br F_{S}$, $\br G_{T}$, and $\br G_{S}$ all
vanish at infinity fast enough to produce uniqueness. Now define $\br W=\br F-\br G$.
A possible decomposition of $\br W$ is then $\br W=\br W_{T}+\br W_{S}$
where $\br W_{T}=\left(\br F_{T}-\br G_{T}\right)$ and $\br W_{S}=\left(\br F_{S}-\br G_{S}\right)$.
Then $\br W_{T}$ and $\br W_{S}$ also vanish at infinity fast enough
to be unique.

Now let $\br F=\br G$ so that $\br W=0$. An obvious possible decomposition
of $\br W=0$ is then $\br W_{T}=0$ and $\br W_{S}=0$. But we have
just shown $\br W_{T}$ and $\br W_{S}$ to be unique. Therefore $\br F=\br G$
implies 
\begin{equation}
0=\br W_{T}=\left(\br F_{T}-\br G_{T}\right)\quad\quad\text{and}\quad\quad0=\br W_{S}=\left(\br F_{S}-\br G_{S}\right)\label{eq:8.7}
\end{equation}
and hence $\br F_{T}=\br G_{T}$ and $\br F_{S}=\br G_{S}$.

If uniquely decomposed functions $\br F$ and $\br G$ are equal,
then their decomposed parts are separately equal; $\br F=\br G$ if
and only if $\br F_{T}=\br G_{T}$ and $\br F_{S}=\br G_{S}$.

\section{Textbook Treatments of Uniqueness\label{sec:Text-Treatments}}

Textbook treatments of the condition for the uniqueness of a vector
function $\br F$ given its divergence and curl vary. Two older texts
overestimate the required rapidness of the vanishing of $\br F$ as
$r\rightarrow\infty$, and two more recent texts appear to underestimate
it.

Stratton \citep{stratton} overestimates the condition for uniqueness
of a vector function given its divergence and curl. On page 196, using
notation defined on page 168, he states the condition for uniqueness
in Green's identity to be (in our notation) $\tilde{\phi}=O(r^{-1})$
and $\nabla\tilde{\phi}=O(r^{-2})$. Comparing this to the condition
$\tilde{\phi}=O(r^{-(1/2+\beta)})$ and $\nabla\tilde{\phi}=O(r^{-(3/2+\beta)})$
derived in our \prettyref{sec:Uniq-whole}, shows Stratton's condition
certainly to be sufficient since $1>1/2$ and $2>3/2$. However, his
condition is unnecessarily strong.

In the text following eqn.(1-21) on page 5, Panofsky and Phillips
\citep{PanofskyPhillips} state that the condition for uniqueness
of a vector field given its divergence and curl to be that their function
$\psi$ in Green's identity, \textquotedbl{}...tends to zero at least
as $1/r$.\textquotedbl{} In our notation, this is $\tilde{\phi}=O(r^{-1})$.
Comparing this to the condition $\tilde{\phi}=O(r^{-(1/2+\beta)})$
derived in our \prettyref{sec:Uniq-whole}, shows this condition certainly
to be sufficient since $1>1/2$. However, as with Stratton, this condition
is unnecessarily strong.

On the other hand, in his discussion of Dirichlet and Neumann boundary
conditions on page 43, Jackson \citep{jackson} says that these conditions
may apply, \textquotedbl{}...on a closed surface (part or all of which
may be at infinity, of course).\textquotedbl{} When applied in a development
such as our Theorem 5.1, this idea of a surface at infinity would
imply uniqueness when $\br F\rightarrow0$ as $r\rightarrow\infty$
since that would make $\br F=0$ on the surface \emph{at infinity.}
But, as pointed out at the ends of our Sections \ref{sec:Uniq-finite}
and \ref{sec:Uni-Poisson-Finite}, there is no such thing as a surface
\emph{at infinity.} Jackson's condition for uniqueness in the whole
of $E^{3}$ would therefore not be sufficient. The vanishing of $\br F$
as $r\rightarrow\infty$ is not a sufficient condition for uniqueness.
As proved in our \prettyref{sec:Uniq-whole}, the weakest sufficient
condition is $\br F=O(r^{-(3/2+\beta)}).$

On page 222, Jackson quotes the same decomposition of $\br F$ into
solenoidal and transverse vectors as that in our \prettyref{sec:Decomposition}.
But he does not discuss the uniqueness of that decomposition.

On page 583 Griffiths \citep{griffiths} makes the statement, \textquotedbl{}...there
is \emph{no} function that has zero divergence and zero curl everywhere
\emph{and} goes to zero at infinity.\textquotedbl{} The correctness
of that statement depends on the interpretation of the phrase, \textquotedbl{}...goes
to zero at infinity.\textquotedbl{} A strict definition would be $\br F=O(r^{-\alpha})$
for some (possibly small) positive constant $\alpha$. If that definition
is used, Theorem 6.1 shows that his statement must be modified to
require $\alpha=3/2+\beta$ for some (possibly small) positive constant
$\beta$. As noted at the end of \prettyref{sec:Uniq-whole} the correct
statement would be: There is no (nonzero) function that has zero divergence,
zero curl, and goes to zero faster than $r^{-3/2}$ as $r\rightarrow\infty.$ 

Like Jackson, Griffiths underestimates the rapidness with which $\br F$
must approach zero as $r\rightarrow\infty$. The vanishing of $\br F$
as $r\rightarrow\infty$ is not a sufficient condition for uniqueness.
As proved in our \prettyref{sec:Uniq-whole}, the weakest sufficient
condition is $\br F=O(r^{-(3/2+\beta)}).$

Griffiths justifies his statement quoted above by reference to his
Section 3.1.5 that discusses Dirichlet boundary conditions. Thus he
seems to assume, as does Jackson, that there is such a thing as a
surface \emph{at infinity} to which these boundary conditions may
be applied. But, as pointed out at the ends of our Sections \ref{sec:Uniq-finite}
and \ref{sec:Uni-Poisson-Finite}, there is no such thing as a surface
\emph{at infinity. }The vanishing of $\br F$ as $r\rightarrow\infty$
is therefore not a sufficient condition for uniqueness.

\section{Summary\label{sec:Summary}}

The calculation of valid solutions to the classical Maxwell Equations
requires close attention to questions of convergence and uniqueness.
This article corrects some imprecise statements in the textbook literature.
It also derives a new sufficient condition for a vector function to
be determined uniquely by its divergence and curl.

\section{Appendix A\label{sec:Appendix-B}}

This Appendix discusses the use of Green's identity to solve the Poisson
equation in the whole of the space $E^{3}$ with no conductors or
other physically imposed boundary surfaces.

Taking the divergence of the quantities $\phi\nabla\psi$ and $\psi\nabla\phi$
gives
\begin{equation}
\nabla\cdot\left(\phi\nabla\psi\right)=\nabla\phi\cdot\nabla\psi+\phi\nabla^{2}\psi\quad\text{and}\quad\nabla\cdot\left(\psi\nabla\phi\right)=\nabla\psi\cdot\nabla\phi+\psi\nabla^{2}\phi\label{eq:13.1}
\end{equation}
Subtracting these two equations gives 
\begin{equation}
\nabla\cdot\left(\phi\nabla\psi-\psi\nabla\phi\right)=\left(\phi\nabla^{2}\psi-\psi\nabla^{2}\phi\right)\label{eq:13.2}
\end{equation}
Using the divergence theorem for a volume ${\cal V}$ with surface
${\cal S}$ then gives Green's identity in the form
\begin{equation}
\oint_{{\cal S}}\left(\phi\nabla\psi-\psi\nabla\phi\right)\cdot d\br a=\int_{{\cal V}}\left(\phi\nabla^{2}\psi-\psi\nabla^{2}\phi\right)d\tau\label{eq:13.3}
\end{equation}

For a linear partial differential operator $\nabla^{2}$, the Green's
function $G(\br r,\br r_{0})$ is a solution of the equation 
\begin{equation}
\nabla^{2}G(\br r,\br r_{0})=\delta^{3}\left(\br r-\br r_{0}\right)\label{eq:13.4}
\end{equation}
 where $\delta^{3}\left(\br r-\br r_{0}\right)$ is the Dirac delta
function. A solution to \prettyref{eq:13.4} is 
\begin{equation}
G(\br r,\br r_{0})=-\dfrac{1}{4\pi}\,\dfrac{1}{\left|\br r-\br r_{0}\right|}\label{eq:13.5}
\end{equation}
When defined, as here, in the whole of $E^{3}$ with no physically
imposed surfaces, the Green's function defined in \prettyref{eq:13.5}
is also called the \emph{fundamental solution} since its convolution
with source term $-f$ gives the solution $U.$

Now apply Green's identity in \prettyref{eq:13.3} with $\phi=U$
and $\psi=G$, where $U$ is a solution to the Poisson equation
\begin{equation}
\nabla^{2}U(\br r)=-f(\br r)\label{eq:13.6}
\end{equation}
The result is
\[
\oint_{{\cal S}}\left[U(\br r)\nabla G(\br r,\br r_{0})-G(\br r,\br r_{0})\nabla U(\br r)\right]\cdot d\br a
\]
\begin{equation}
=\int_{{\cal V}}\left[U(\br r)\nabla^{2}G(\br r,\br r_{0})-G(\br r,\br r_{0})\nabla^{2}U(\br r)\right]d\tau\label{eq:13.7}
\end{equation}
\[
=\int_{{\cal V}}U(\br r)\delta^{3}\left(\br r-\br r_{0}\right)d\tau-\dfrac{1}{4\pi}\int_{{\cal V}}\frac{f(\br r)}{\left|\br r-\br r_{0}\right|}d\tau
\]
Evaluating the delta function term, the result is 
\begin{equation}
U(\br r_{0})=\dfrac{1}{4\pi}\int_{{\cal V}}\frac{f(\br r)}{\left|\br r-\br r_{0}\right|}d\tau-\oint_{{\cal S}}\left[U(\br r)\nabla G(\br r,\br r_{0})-G(\br r,\br r_{0})\nabla U(\br r)\right]\cdot d\br a\label{eq:13.8}
\end{equation}
A trial solution is 
\begin{equation}
U(\br r_{0})=\dfrac{1}{4\pi}\int_{{\cal V}}\frac{f(\br r)}{\left|\br r-\br r_{0}\right|}d\tau\label{eq:13.9}
\end{equation}

To show this trial solution correct, evaluate the surface integral
in \prettyref{eq:13.8} on a spherical surface of radius $\rho$ centered
at the origin. For trial solution \prettyref{eq:13.9} to be correct,
that surface integral must vanish as $\rho\rightarrow\infty$ and
${\cal V}$ becomes the whole of the space. 

The condition $f=O(r^{-2-\varepsilon})$ was shown in \prettyref{sec:Solution-Poisson}
to be necessary in order for the integrand in \prettyref{eq:13.9}
to be $O(r^{-1-\varepsilon})$ and hence for that integral to converge.
It follows that $U=O(r^{-\varepsilon})$ and $\nabla U=O(r^{-1-\varepsilon})$.
Also $G=O(r^{-1})$ and $\nabla G=O(r^{-2})$. Since $da=r^{2}d\Omega$,
the surface integral in \prettyref{eq:13.8} is $O(r^{-1-\varepsilon-1+2})=O(r^{-\varepsilon})$
and hence goes to zero as $\rho\rightarrow\infty$, as required. When
$f=O(r^{-2-\varepsilon})$, the trial solution \prettyref{eq:13.9}
is correct.

\end{document}